\documentclass[%
 aip,
 amsmath,amssymb,
 reprint,%
]{revtex4-1}

\usepackage{graphicx}
\usepackage{dcolumn}
\usepackage{bm}

\usepackage[utf8]{inputenc}
\usepackage[T1]{fontenc}
\usepackage{mathptmx}
\usepackage{etoolbox}

\usepackage[separate-uncertainty = true]{siunitx}
\DeclareSIUnit\torr{Torr}

\usepackage[final]{changes}
\usepackage{makecell}

\usepackage[english]{babel}

\makeatletter
\def\@email#1#2{%
 \endgroup
 \patchcmd{\titleblock@produce}
  {\frontmatter@RRAPformat}
  {\frontmatter@RRAPformat{\produce@RRAP{*#1\href{mailto:#2}{#2}}}\frontmatter@RRAPformat}
  {}{}
}%
\makeatother
\begin{document}

\preprint{AIP/123-QED}

\title[Tungsten Germanide SNSPDs with Saturated Internal Detection Efficiency at Wavelengths up to 29 µm]{Tungsten Germanide Superconducting Nanowire Single-Photon Detectors with Saturated Internal Detection Efficiency at Wavelengths up to 29 µm}
\author{Benedikt Hampel}
\email{benedikt.hampel@nist.gov}
\thanks{These authors contributed equally.}
\affiliation{National Institute of Standards and Technology, Boulder, Colorado 80305, USA}
\affiliation{Department of Physics, University of Colorado Boulder, Boulder, Colorado 80309, USA}

\author{Daniel Kuznesof}
\thanks{These authors contributed equally.}
\affiliation{James Watt School of Engineering, University of Glasgow, Glasgow G12 8QQ, United Kingdom}

\author{Andrew~S.~Mueller}
\affiliation{Jet Propulsion Laboratory, California Institute of Technology, Pasadena, California 91109, USA}

\author{Sahil R. Patel}
\affiliation{Jet Propulsion Laboratory, California Institute of Technology, Pasadena, California 91109, USA}

\author{Robert H. Hadfield}
\affiliation{James Watt School of Engineering, University of Glasgow, Glasgow G12 8QQ, United Kingdom}

\author{Emma~E.~Wollman}
\affiliation{Jet Propulsion Laboratory, California Institute of Technology, Pasadena, California 91109, USA}

\author{Matthew D. Shaw}
\affiliation{Jet Propulsion Laboratory, California Institute of Technology, Pasadena, California 91109, USA}

\author{Dirk Schwarzer}
\affiliation{Max Planck Institute for Multidisciplinary Sciences, Department of Dynamics at Surfaces, 37077 Göttingen, Germany}

\author{Alec M. Wodtke}
\affiliation{Max Planck Institute for Multidisciplinary Sciences, Department of Dynamics at Surfaces, 37077 Göttingen, Germany}

\author{Khalid Hossain}
\affiliation{JP Analytical, Richardson, Texas 75081, USA}

\author{Allison V. Mis}
\affiliation{National Institute of Standards and Technology, Boulder, Colorado 80305, USA}

\author{Alexana~Roshko}
\affiliation{National Institute of Standards and Technology, Boulder, Colorado 80305, USA}

\author{Richard P. Mirin}
\affiliation{National Institute of Standards and Technology, Boulder, Colorado 80305, USA}

\author{Sae Woo Nam}
\affiliation{National Institute of Standards and Technology, Boulder, Colorado 80305, USA}

\author{Martin J. Stevens}
\affiliation{National Institute of Standards and Technology, Boulder, Colorado 80305, USA}

\author{Varun B. Verma}
\thanks{These authors contributed equally.}
\affiliation{National Institute of Standards and Technology, Boulder, Colorado 80305, USA}


\date{\today}

\begin{abstract}
Superconducting nanowire single-photon detectors (SNSPDs) are among the most sensitive single-photon detectors available and have the potential to transform fields ranging from infrared astrophysics to molecular spectroscopy. However, extending their performance into the mid-infrared spectral region - crucial for applications such as exoplanet transit spectroscopy and vibrational fingerprinting of molecules - has remained a major challenge, primarily due to material limitations and scalability constraints. Here, we report on the development of SNSPDs based on tungsten germanide, a novel material system that combines high mid-infrared sensitivity with compatibility for large-scale fabrication. Our detectors exhibit saturated internal detection efficiency at wavelengths up to \qty{29}{\micro\meter}, while using \num[]{2.7}$\times$ thicker films (\qty{8}{\nano\meter} vs \qty{3}{\nano\meter}) and up to \num[]{4.5}$\times$ wider nanowires (\qty{360}{\nano\meter} vs \qty{80}{\nano\meter}) compared to mid-infrared-optimized SNSPDs fabricated from tungsten silicide. This advance will enable scalable, high-performance single-photon detection in a spectral region that was previously inaccessible, opening new frontiers in remote sensing, thermal imaging, environmental monitoring, molecular physics, and astronomy.
\end{abstract}

\maketitle

\section{Introduction}
Mid-infrared (mid-IR) single photon detection is an enabling technology for applications in areas like physical chemistry \cite{Lau.2023}, astronomy \cite{Wollman.2024, Glauser.2024}, and commercial sensing \cite{taylor_photon_2019}. Such applications include observations of infrared chemiluminescence \cite{Polanyi.1963}, mid-IR laser-induced fluorescence \cite{Stewart.1983} and atmospheric emissions \cite{Mlynczak.1997}, spectroscopic techniques in the single photon limit \cite{wollman_recent_2021,Lau.2023}, optical coherence tomography for non-destructive penetrative imaging \cite{israelsen_real-time_2019, fang_mid-infrared_2023}, as well as photonic quantum computing with spin qubits and quantum communications \cite{yan_silicon_2021,li_quantum_2024, temporao_feasibility_2008}.  Mid-IR detectors are also prized in the fields of medicine and biology, where single photon sensitivity would enable faster acquisition times, greater dynamic range \cite{yeh_infrared_2023}, and can exploit the specificity of a biological molecule's unique spectrum \cite{adams_midinfrared_2001} accessible only in the mid-IR.

The dominant commercially available mid-IR detector technology is based on mercury cadmium telluride (MCT or HgCdTe) \cite{lei_progress_2015,hadfield_single-photon_2023}. However, MCT detectors typically have poor single-photon sensitivity and significant noise from readout and dark current. Moreover, they are in the process of being replaced by alternative mid-IR semiconductor-based technologies due to health and environmental concerns. Alternative mid-infrared detectors include Type II superlattice InAs/GaSb which are operated at \qty{77}{\kelvin} but have a shorter spectral range than HgCdTe-based detectors\cite{rogalski_inasgasb_2017}. Silicon-based blocked impurity band (BIB) \cite{xiao_progress_2022} detectors also known as impurity band conduction (IBC) detectors possess sensitivity across a large spectral range (typically \qtyrange[]{2}{28}{\micro\meter} wavelength range) and high quantum efficiency, yet do not possess single-photon sensitivity and must be operated at temperatures of less than \qty{10}{\kelvin}. \deleted{Furthermore, low-noise IBC-based detectors have not been in production for a decade, and it is uncertain whether a path exists to restarting this capability\cite{Roellig}.}

Alternatively, superconducting nanowire single-photon detectors (SNSPDs) have demonstrated single-photon detection up to a wavelength of \qty{29}{\micro\meter} \cite{taylor_low-noise_2023}, near unity system detection efficiency at \qty{1550}{\nano\meter} wavelength \cite{Reddy.2020}, and less than \qty{3}{\pico\second} of timing jitter\cite{Korzh.2020}. Recent developments in system architecture have led to growing maturity of multiplexing schemes such as the thermally-coupled imager \cite{McCaughan.2022}. This multiplexing scheme was used to create a  \num{400000}-pixel SNSPD array \cite{Oripov.2023}, and was utilized for the development of a \num[]{64}-pixel mid-infrared imager array \cite{hampel_64-pixel_2024}.

The materials development of SNSPDs for the mid-IR has been focused on tungsten silicide (WSi). The WSi films are engineered for high resistivity and low critical temperature through co-sputtering silicon (Si) and tungsten (W) \cite{Verma.2021} at a high silicon fraction in order to lower the characteristic energy \textit{E}\textsubscript{0} (Eq.~(\ref{characteristic_energy_eqn})) of the material which scales with the single photon detection energy threshold of SNSPDs. This is dependent on the electron density of states \textit{N(0)} (Eq.~(\ref{density_of_states_eqn})), Boltzman's constant \textit{k}\textsubscript{b}, the critical temperature \textit{T}\textsubscript{c}, and the characteristic volume \textit{V}\textsubscript{0} \cite{Vodolazov.2017}.

\begin{eqnarray}\label{characteristic_energy_eqn}
E_\text{0} = 4 N(0) (k_\text{B}T_\text{c})^2 V_\text{0}
\end{eqnarray}

The density of states depends on the normal state resistivity $\rho$, the electron charge $e$, and the diffusivity $D$ of the material \cite{Vodolazov.2017}. 

\begin{eqnarray}\label{density_of_states_eqn}
N(0) = (2\rho e^2D)^{-1}
\end{eqnarray}

Making high-resistivity WSi films through increasing the Si content has been very successful at extending the cut-off wavelength of SNSPDs \cite{Verma.2021, taylor_low-noise_2023,colangelo2022large}. Furthermore, reducing the nanowire width and thickness decreases the volume and reduces the critical temperature of the film further decreasing the characteristic energy and increasing the cut-off wavelength. However, \replaced{one potential disadvantage of the use of WSi in the mid-infrared is the presence of strong phonon absorption in both the silicon capping layer used to protect the WSi superconducting film as well as in the WSi film itself}{ calibrated efficiency measurements at \hbox{\qty{10}{\micro\meter}} on Si-rich WSi devices fabricated upon dielectric optical stacks with simulated system detection efficiencies in excess of \hbox{\qty{60}{\percent}} have shown measured system detection efficiencies of less than \hbox{\qty{1}{\percent}}. De-convolution of the possible causes of this is challenging. This unexpectedly poor performance could be due to the high silicon content of the WSi; silicon may strongly absorb mid-infrared photons leading to phonon excitation} \cite{Wollack.2020}. The energy associated with these phonons may then leave the film through the substrate before interacting with the electron system, thus preventing the breaking of Cooper-pairs which would lead to a detection event. \deleted{Additionally, the \hbox{\qty{2}{\nano\meter}}-thick protective amorphous silicon capping layer on top of these SNSPDs could be contributing to photon absorption outside of the detector region. }These photons could therefore not contribute to detection events resulting in a significantly diminished system detection efficiency.

\replaced{Since germanium is chemically very similar to silicon and also has}{ These hypotheses regarding the origin of poor system detection efficiency in WSi instigated an investigation of germanium as a possible replacement for silicon. With} good transmissive properties for wavelengths up to \qty{14}{\micro\meter}, \replaced{WGe is naturally an interesting alternative to WSi to}{it is a promising candidate to } investigate for high-performance mid-IR SNSPDs. In the literature \cite{ercolano_magnetoconductivity_2024}, WGe has also been highlighted as being particularly suited for mid-IR single photon detection due to the electronic system efficiently retaining the energy of absorbed photons, enhancing WGe detectivity at low photon energies. This idea can be quantified by the ratio of the electron-phonon ($\tau$\textsubscript{e-ph}) interaction timescale and the electron-electron timescale ($\tau$\textsubscript{e-e}). Additionally, this metric has been reported to be marginally larger in WGe (\num[]{2.6 +- 0.5}) than in WSi (\num[]{2.5 +- 0.5}) \cite{ercolano_magnetoconductivity_2024}. SNSPDs based on germanides in the past have been demonstrated such as MoGe and WGe. However, single photon sensitivity has only been showcased for wavelengths of up to \qty{2}{\micro\meter} \cite{verma_superconducting_2014,Yang.2025}. 

Here, we enhance the mid-infrared sensitivity and extend the maximum cut-off wavelength through the further optimization of tungsten germanide SNSPDs for mid-infrared applications beyond \qty{10}{\micro\meter} wavelength. This material platform enables greater compatibility with Ge-based dielectric optical cavities, wider nanowires (\qtyrange[]{100}{360}{\nano\meter}), and thicker films (\qty{8}{\nano\meter}) allowing for the possibility of high efficiency single photon detectors in the mid-infrared and improved scalability for the fabrication of imaging arrays.

\section{Fabrication}
WGe SNSPDs were fabricated from germanium-rich WGe films for high resistivity and suppressed critical temperature. Multiple bulk films were co-sputtered from separate tungsten and germanium targets using dc sputtering at a pressure of \qty{3}{\milli\torr} and with the substrate at room temperature. The films were created by sputtering for \qty{5}{minutes}, resulting in a thickness of approximately \qty{50}{\nano\meter}. The germanium target sputtering power was fixed at \qty{60}{\watt}, while the tungsten target sputtering power was varied from \qtyrange[]{10}{200}{\watt}. \added{Note that the film thickness of the different samples varies slightly with the change of tungsten sputtering power. The sputtering rate was only determined precisely for the sample with \hbox{\qty{20}{\watt}} sputtering power of the tungsten target. }The films were capped in-situ by a protective amorphous germanium layer with a thickness of approximately \qty{5}{\nano\meter}.

\begin{figure}[htbp]
    \centering
    \includegraphics[width = 0.45\textwidth]{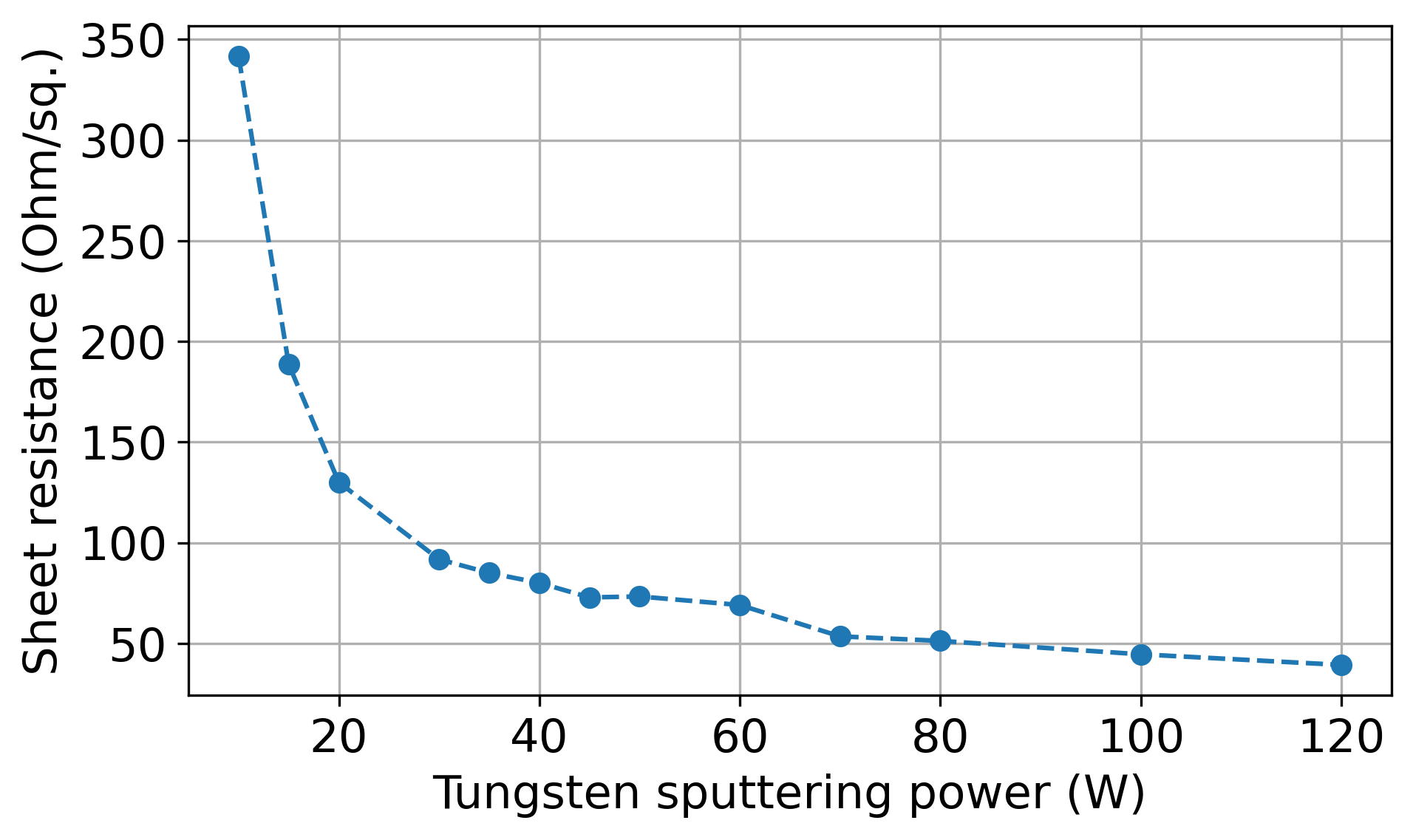}
    \caption{\label{Rsq_vs_Wpower}Four-probe room temperature sheet resistance of co-sputtered bulk WGe films as a function of tungsten target sputtering power.}
\end{figure}

Four-probe resistivity measurements were made at room temperature to determine the sheet resistance of samples as depicted in Fig.~\ref{Rsq_vs_Wpower}. It can be seen that the sheet resistance is exponentially growing for lower tungsten sputtering powers. This is desired for the development of single-photon detectors with high detection efficiencies as this can be correlated with a lower density of charge carriers. The energy of an absorbed photon would therefore be distributed amongst fewer Cooper pairs in the superconductor, which lowers the energy threshold required to generate a detection event since the average temperature of the electron system post-absorption is then higher. 

\begin{figure}[htbp]
    \centering
    \includegraphics[width = 0.45\textwidth]{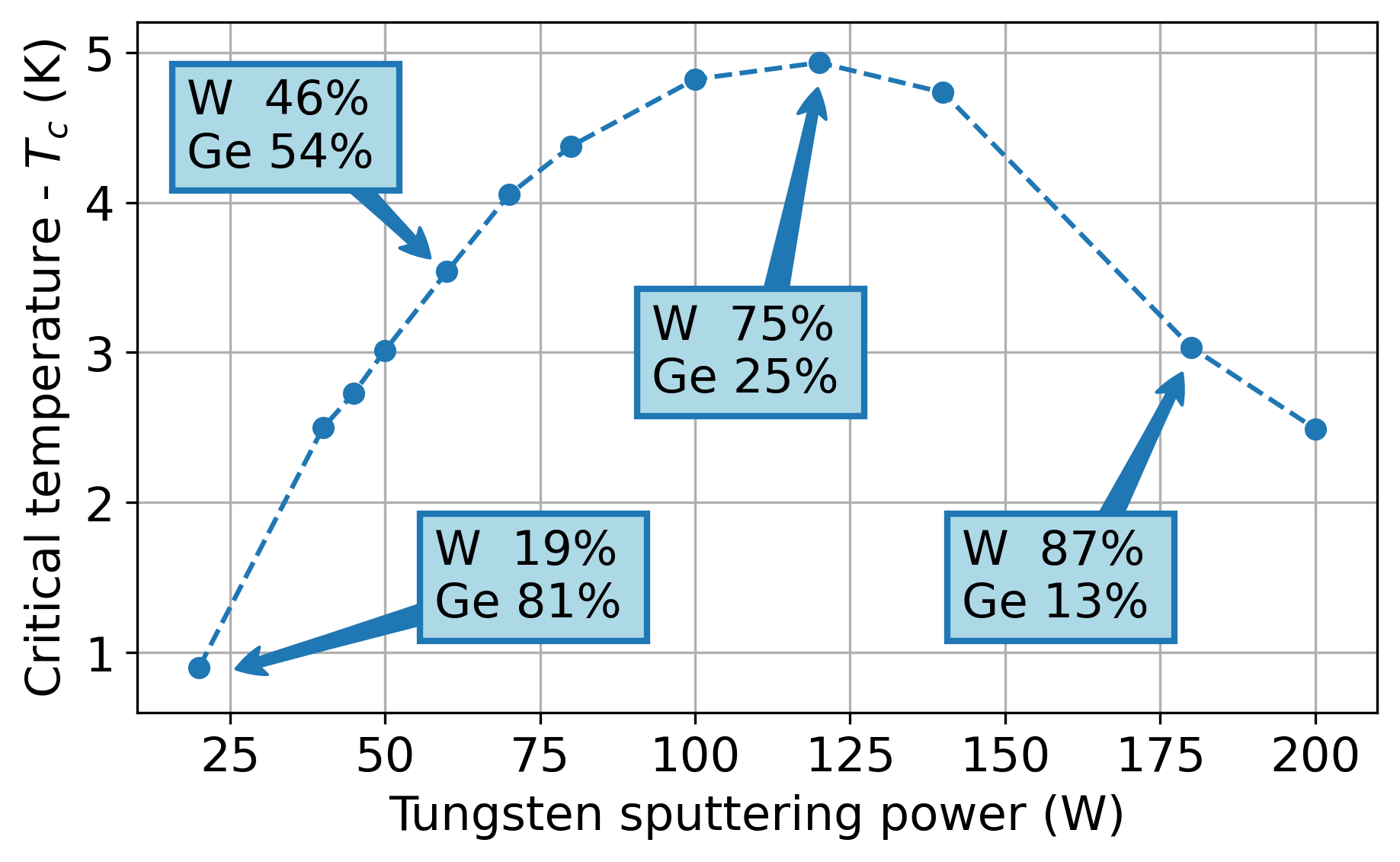}
    \caption{\label{Tc_W_target_sputtering_power}Superconducting transition temperature \textit{T}\textsubscript{c} as a function of sputtering power of the tungsten target during co-sputtering of the WGe. The germanium sputtering power was fixed at \qty{60}{\watt}. Measurements were performed on bulk films with thickness of approximately \qty{50}{\nano\meter}. The insets show the tungsten (W) and germanium (Ge) film composition of some samples, which was determined by Rutherford backscattering spectrometry (RBS).}
\end{figure}

The critical temperature of the films was determined by four-probe resistivity measurements in a cryostat with a base temperature of about \qty{2}{\kelvin}. The resistance was measured during cooling and warmup at a rate of \qty{1}{\kelvin\per\minute} to avoid thermal hysteresis effects. The results for samples fabricated with different tungsten sputtering powers are depicted in Fig.~\ref{Tc_W_target_sputtering_power}. The composition of four samples was analyzed by Rutherford backscattering spectrometry (RBS). The results can be seen in the insets of Fig.~\ref{Tc_W_target_sputtering_power}. 

A tungsten target sputtering power of \qty{20}{\watt} was chosen for the deposition of thin films suitable for the fabrication of mid-IR-optimized SNSPDs based on past experience with WSi, and the effects of suppressed \textit{T}\textsubscript{c} and higher sheet resistivity. These material properties enable a much smaller characteristic energy \textit{E}\textsubscript{0}, which makes the thin film more suitable for mid-IR single photon detection as shown in Eq.~(\ref{characteristic_energy_eqn}) and Eq.~(\ref{density_of_states_eqn}). The \textit{T}\textsubscript{c} of this film was below the base temperature of the previously used cryostat so that it was measured in an adiabatic demagnetization refrigerator with a base temperature of \qty{180}{\milli\kelvin} using a two-probe resistivity measurement. The \textit{T}\textsubscript{c} was determined to be \qty{900}{\milli\kelvin}.  

Fabrication of SNSPDs was performed on a 3-inch oxidized silicon wafer with \qty{150}{\nano\meter} of thermally-grown SiO\textsubscript{2}. The \qty{8}{\nano\meter}-thick WGe thin film was then sputtered with a tungsten sputtering power of \qty{20}{\watt}, and the film was capped in-situ with a 4 nm-thick layer of amorphous germanium. Gold bonding pads and alignment marks for electron beam and optical lithography were then deposited by electron beam evaporation and liftoff. Electron beam lithography was performed to define the SNSPD patterns using \qty{200}{\nano\meter}-thick polymethyl methacrylate (PMMA) resist. Multiple SNSPDs were defined on each chip having wire widths between \qtyrange[]{100}{360}{\nano\meter} with a fill factor of \qty{30}{\percent} to minimize current crowding \cite{clem_geometry-dependent_2011}. The active area of each SNSPD was designed to be \qtyproduct{16 x 16}{\micro\meter}. Reactive ion etching (RIE) in an SF\textsubscript{6} plasma was used to transfer the pattern from PMMA into the underlying WGe film. The PMMA was then stripped in acetone, and a \qty{20}{\micro\meter}-wide stripe was patterned by optical lithography connecting the bond pads to ground and overlapping with the SNSPD to remove the WGe film on the remainder of the wafer. The remaining WGe was again etched by RIE in an SF\textsubscript{6} plasma, and the photoresist stripped in acetone. Finally, a \qty{15}{\nano\meter}-thick layer of germanium was sputtered over the entire wafer to encapsulate the SNSPD and protect it from oxidation. 

\begin{figure*}[ht]
    \centering\includegraphics[width = 1.0\textwidth]{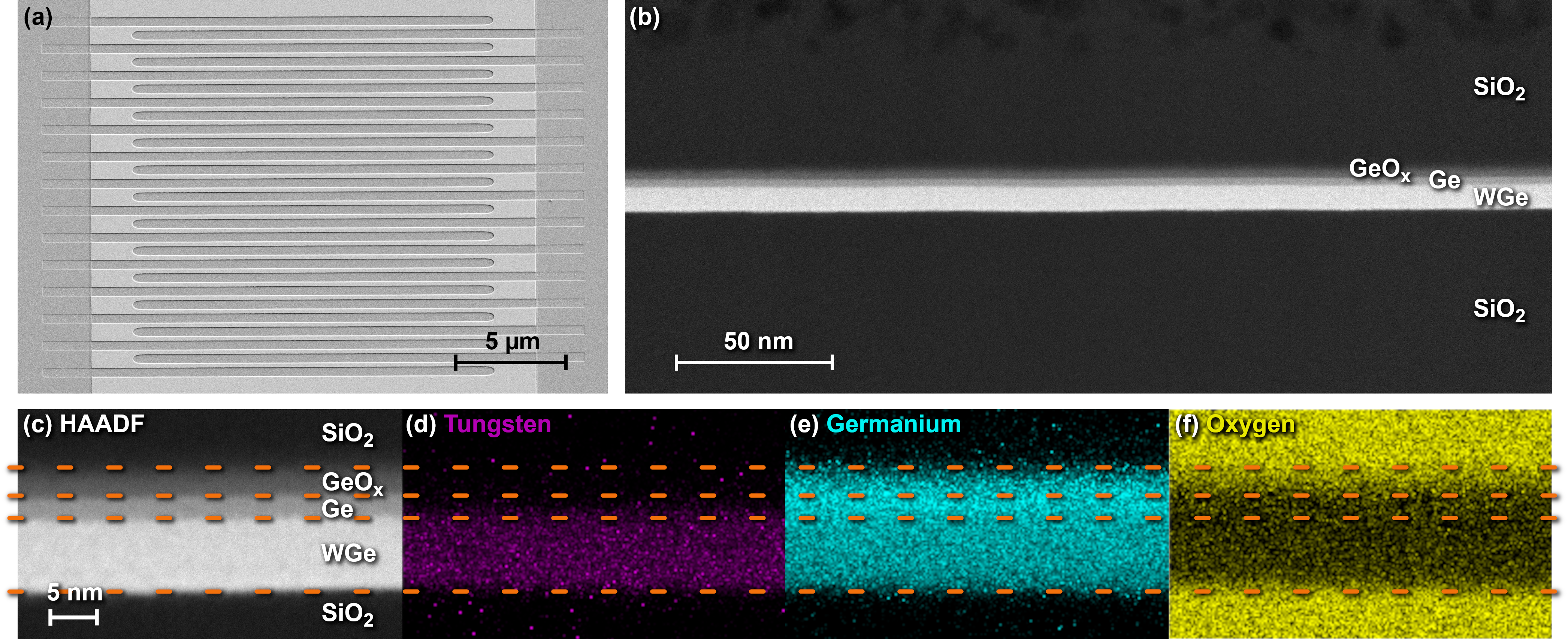}
    \caption{\label{SNSPD_SEM_TEM}(a) Scanning electron micrograph of an SNSPD meander with \qty{200}{\nano\meter} wide nanowires on a \qty{600}{\nano\meter} pitch. The size of the active area is \qtyproduct{16 x 16}{\micro\meter}. (b) High-angle annular dark-field scanning transmission electron micrograph (HAADF-STEM) of the layer stack. (c) HAADF-STEM micrograph of the WGe layer with the Ge capping layer in comparison to energy dispersive X-ray spectroscopy (EDS) analyses showing (d) the tungsten signals, (e) the germanium signals, and (f) the oxygen signals. The dashed orange lines serve as guides to the eye to illustrate the layer structure.}
\end{figure*}

The scanning electron micrograph (SEM) in Fig.~\ref{SNSPD_SEM_TEM}(a) shows a patterned device after the entire process has been completed. The layer stack was analyzed with a scanning transmission electron microscope (STEM) which is equipped with a high-angle annular dark-field (HAADF) detector and an energy dispersive X-ray spectroscopy (EDS) detector. A very good uniformity of the layer stack of WGe and the Ge cap can be seen in Fig.~\ref{SNSPD_SEM_TEM}(b). A \qty{150}{\nano\meter}-thick SiO\textsubscript{2} protection layer was used instead of the \qty{15}{\nano\meter} Ge protection layer to measure the thickness of the Ge capping layer accurately. The dark patterns in the \added{top region of the }top SiO\textsubscript{2} layer \added{in Fig.~\ref{SNSPD_SEM_TEM}(b)} originate from the TEM sample preparation with a focused ion beam (FIB). The thickness of the WGe layer was determined to be \qty{7.7}{\nano\meter} with a \qty{5.1}{\nano\meter} germanium cap of which \qty{2.7}{\nano\meter} are oxidized. Figure~\ref{SNSPD_SEM_TEM}(c) shows a detailed HAADF micrograph of the WGe, Ge and GeO\textsubscript{x} layers. Its contrast aligns well with the separation in the EDS maps for tungsten in Fig.~\ref{SNSPD_SEM_TEM}(d), for germanium in Fig.~\ref{SNSPD_SEM_TEM}(e), and for oxygen in Fig.~\ref{SNSPD_SEM_TEM}(f).  The increased oxygen content in the top \qty{2.7}{\nano\meter} of the germanium cap demonstrates the extent of oxidation due to exposure to ambient air during processing.

\section{Measurement Setup}

\begin{figure}[ht]
    \centering\includegraphics[width = 0.45\textwidth]{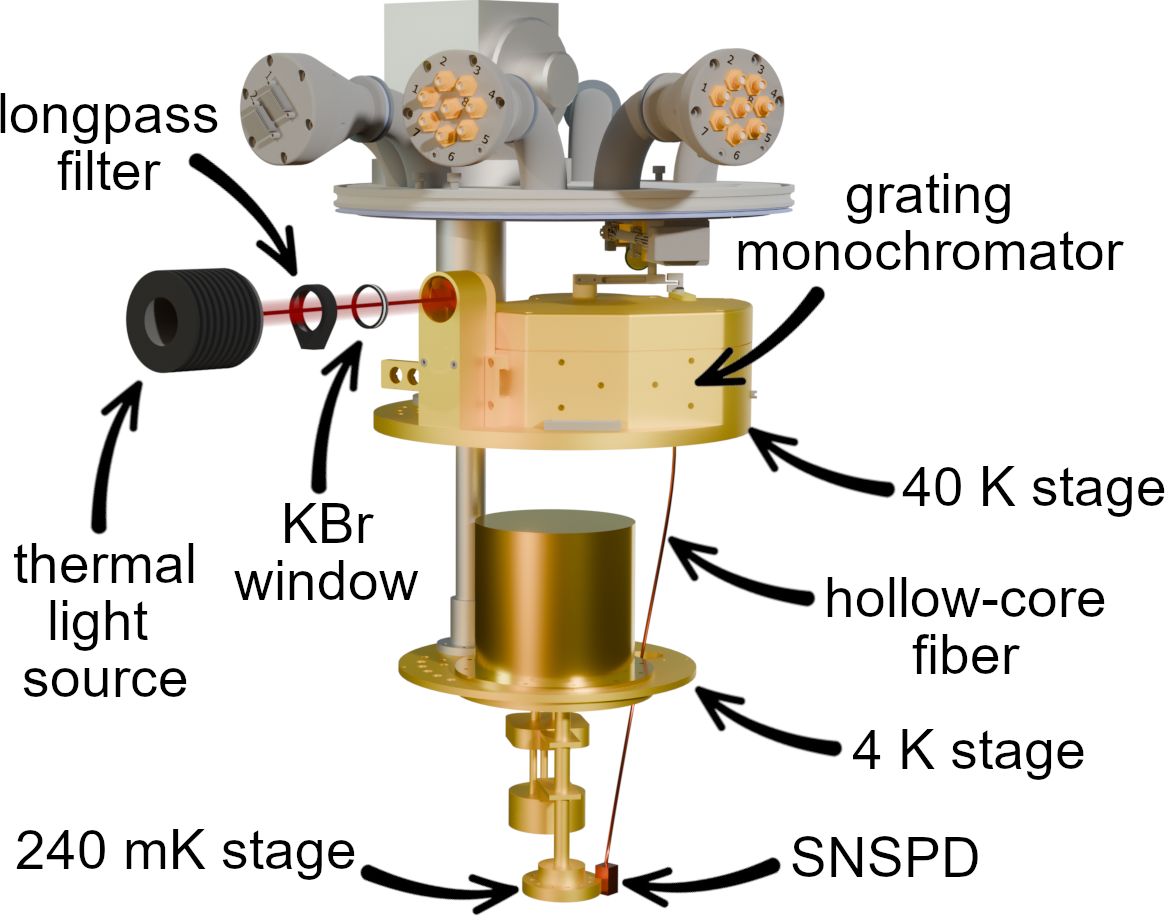}
    \caption{\label{setup}Scanning grating monochromator cryostat system based upon a blazed diffractive grating. The thermal shields at the \qty{40}{\kelvin} stage and the \qty{4}{\kelvin} stage together with the vacuum chamber are removed in the model to show the inside of the system.}
\end{figure}

The SNSPDs were housed inside a cryostat that combined a Gifford-McMahon cryocooler with a \textsuperscript{3}He sorption stage cooler with a base temperature of \qty{240}{\milli\kelvin}. The setup is depicted in Fig.~\ref{setup}\added{ and is different from the setups that were used for the \textit{T}\textsubscript{c} measurements}. A broadband blackbody source with a temperature of \qty{630}{\kelvin} is used as thermal light source. The light passes through a longpass filter and is free-space-coupled through a \qty{5}{\milli\meter}-thick potassium bromide (KBr) optical window into the cryostat. A scanning grating monochromator at the \qty{40}{\kelvin} stage uses a blazed diffraction grating \cite{chen_mid-infrared_2017, Lau.2020, Lau.2023} that allows the precise selection of mid-IR wavelengths with a bandwidth of about \qty{30}{\nano\meter}. The photons are coupled into a hollow-core fiber with a \qty{200}{\micro\meter} core diameter with over \qty{80}{\percent} transmission for wavelengths between \qtyrange{3}{10}{\micro\meter}. The fiber ends in the housing of the SNSPD at the \qty{240}{\milli\kelvin} stage to flood illuminate the whole chip from a distance of \qty{2}{\centi\meter}. 

The wavelength-dependent photo response of each SNSPD at \qty{4}{\micro\meter}, \qty{6}{\micro\meter}, \qty{8}{\micro\meter}, and \qty{10}{\micro\meter} wavelength was investigated. A \qty{3.6}{\micro\meter} long pass filter was used for the measurements at \qty{4}{\micro\meter}, and a \qty{4.5}{\micro\meter} long pass filter was used for measurements at \qty{6}{\micro\meter}. Measurements at \qty{8}{\micro\meter} and \qty{10}{\micro\meter} were performed together with a \qty{7.3}{\micro\meter} long pass filter. These long pass filters removed higher order diffraction components that the grating spectrometer would otherwise let pass from the external broadband blackbody source.   

The SNSPDs were biased through a bias-tee that was integrated into a cryogenic amplifier. The signal from the SNSPDs was amplified by this amplifier that is attached to the \qty{4}{\kelvin}-stage. It has a gain of \qty{45}{\decibel}, a bandwidth of \qty{1.5}{\giga\hertz}, and a noise temperature of less than \qty{6}{\kelvin}. An additional room-temperature amplifier with a gain of \qty{25}{\decibel} and a bandwidth of \qty{500}{\mega\hertz} was used. This provides a high signal-to-noise ratio and enabled pulses to be read out at bias currents of less than \qty{250}{\nano\ampere}. A cryogenic shunt comprised of a \qty{20}{\ohm} resistor and a \qty{450}{\nano\henry} inductor in series was connected in parallel to the SNSPD at the \qty{240}{\milli\kelvin} stage. This was designed and implemented to improve the operational stability of the devices and to prevent latching to the resistive normal state \cite{Kerman2006,Annunziata2010,Liu2012}. The SNSPD pulses were recorded by a time tagger at a threshold level of \qty{130}{\milli\volt}. 

\begin{figure*}[htbp]
    \centering
    \includegraphics[width = 0.9\textwidth]{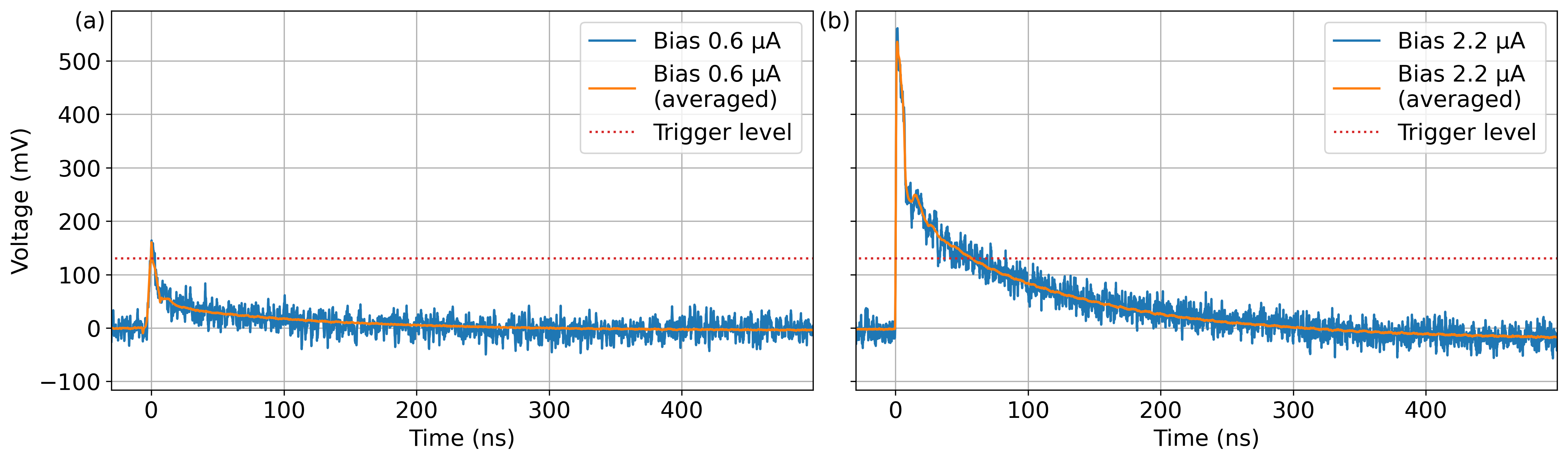}
    \caption{\label{Osci_360nm_130mV}Oscilloscope traces of SNSPD pulses for a bias current of (a) \qty{0.6}{\micro\ampere} and (b) \qty{2.2}{\micro\ampere} of a \qty{360}{\nano\meter} wide and about \qty{8}{\nano\meter} thick nanowire. The orange trace was averaged \qty{1000}{times}. The trigger level for the photon count rate measurements is shown as a red dashed line. }
\end{figure*}

Voltage pulses originating from detection events of a \qty{360}{\nano\meter}-wide SNSPD recorded by a fast oscilloscope at two different bias currents are depicted in Fig.~\ref{Osci_360nm_130mV}. The trigger level for the time tagger is shown as a red dashed line. The decay time of the pulses is very long due to the high kinetic inductance of the WGe film and the shunt. At larger bias currents, the very sensitive time tagger will count multiple detection events due to noise on the decaying signal. The reason for this can be seen in Fig.~\ref{Osci_360nm_130mV}(b) where the threshold is crossed multiple times. To avoid these multiple false counts, an artificial hold-off time of \qty{500}{\nano\second} was introduced in the time tagger software. Any detection events in this time slot after a detection will not be counted as a photon detection event. This will reduce the maximum count rate of the device, but will not impair the characterization of these detectors at relatively low count rates.

A second cryogenic setup was used to characterize WGe SNSPDs at longer wavelengths of \qty{10}{\micro\meter}, \qty{15}{\micro\meter}, \qty{18}{\micro\meter}, and \qty{29}{\micro\meter}. The setup is described in detail in Ref.~\cite{taylor_low-noise_2023}. It employs a thermal light source with a bandpass filter stack inside of the vacuum chamber to flood-illuminate the SNSPDs with mid-infrared photons directly. The electrical part of the setup is similar to the previously described setup. The base temperature of this cryostat is \qty{250}{\milli\kelvin}.

\section{Results}

\begin{figure}[htbp]
    \centering
    \includegraphics[width = 0.45\textwidth]{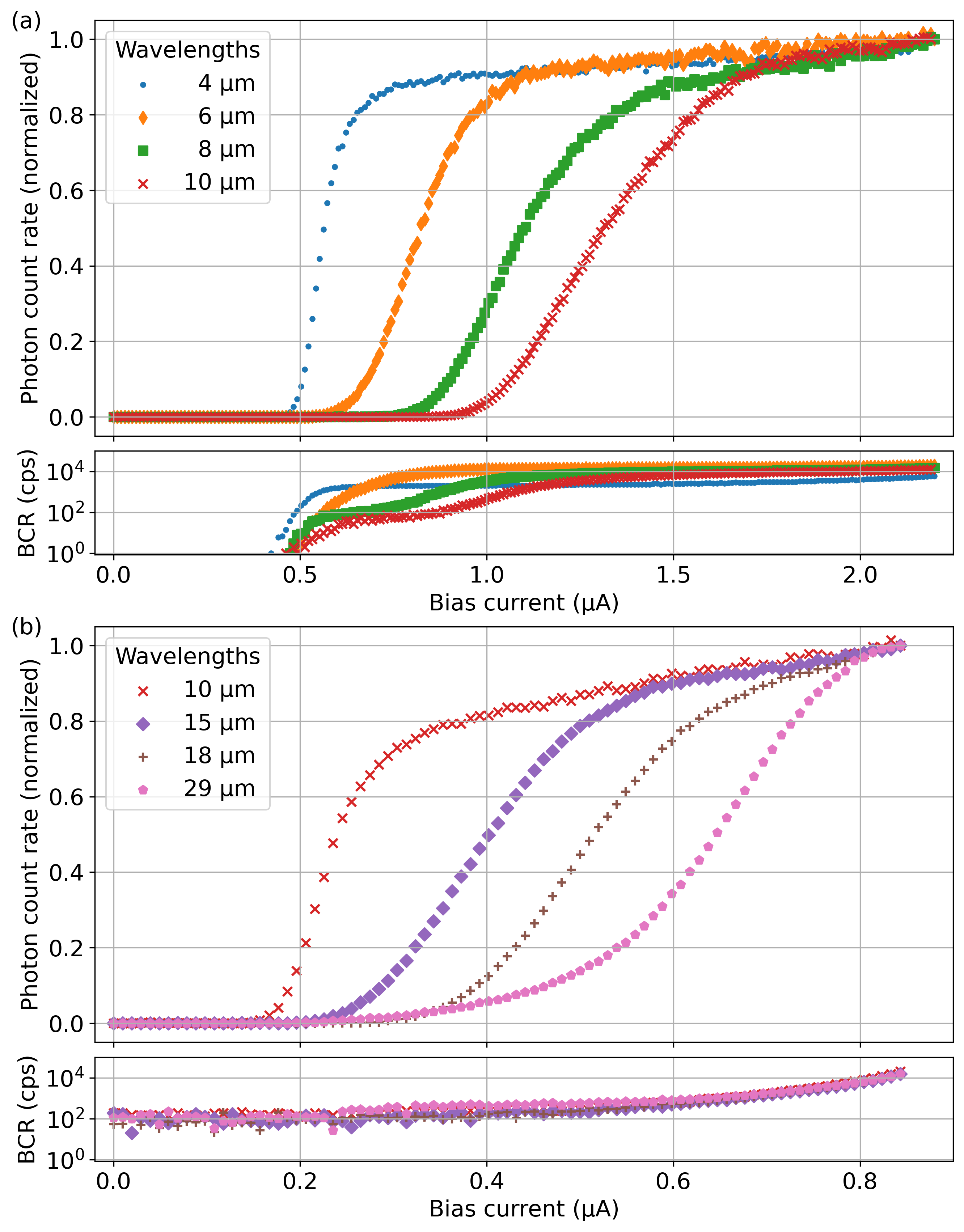}
    \caption{\label{PCR_360nm_130mV}Normalized photon count rate and background count rate (BCR) in counts per second (cps) as a function of bias current of (a) a nanowire with a width of \qty{360}{\nano\meter} for wavelengths between \qtyrange[]{4}{10}{\micro\meter} and of (b) a nanowire with a width of \qty{200}{\nano\meter} for wavelengths between \qtyrange[]{10}{29}{\micro\meter}. The WGe was sputtered at \qty{20}{\watt} and \qty{60}{\watt} for the tungsten and germanium, respectively. The measured BCR is subtracted from the photon count rate measurement and the photon count rate is normalized to the photon count rate value for the highest plotted bias current, \added{which was at about \hbox{\qty{50000}{cps}} for all measurements except for the \hbox{\qty{10}{\micro\meter}} and \hbox{\qty{29}{\micro\meter}} measurements in (b), which were normalized to about \hbox{\qty{26000}{cps}} and \hbox{\qty{62000}{cps}}, respectively.}}
\end{figure}

Figure~\ref{PCR_360nm_130mV}(a) shows the normalized photon count rate as a function of the bias current for wavelengths between \qtyrange[]{4}{10}{\micro\meter} for a \qty{360}{\nano\meter}-wide nanowire meander measured in the cryostat that is equipped with a grating monochromator. The saturated count rate indicates unity internal detection efficiency and suitability of the nanowire for efficient operation at that wavelength \cite{kozorezov_fano_2017}. The slight slope in the plateau regime is likely due to photon absorption and detection in the bends of the meander which are \qty{2}{\micro\meter} wide. This effect could be mitigated by shielding these regions from photon flux.

The background count rate (BCR) is depicted for the different wavelengths in the bottom part of Fig.~\ref{PCR_360nm_130mV}(a). While the measurement setup allows fine tuning of the input wavelength thanks to the monochromator, both the fiber and grating itself are sources of blackbody radiation at mid-IR wavelengths, although this is minimized by cooling the grating to \qty{40}{\kelvin}. \replaced{The dominant source of background photons is blackbody radiation from the room-temperature optical window and longpass filter. These components act effectively as a source even with the blackbody radiation source turned off. In Fig.~\ref{PCR_360nm_130mV}(a), note that measurements of the BCR at \hbox{\qty{8}{\micro\meter}} and \hbox{\qty{10}{\micro\meter}} exhibit a double-plateau structure, with a distinct plateau at low bias currents followed by another plateau at higher bias currents. The plateau at low bias currents is due to second order diffraction from the grating of blackbody radiation emitted by the optical window and longpass filter. The second plateau at high bias currents is due to first order diffraction from the grating of this blackbody radiation. The first plateau is absent at shorter wavelengths due to the low intensity of room-temperature blackbody radiation at second order for those wavelengths.}{ The free-space coupling of the broadband blackbody source allows additional background radiation to enter the cryostat. Higher energy photons, which are not filtered by the longpass filter between source and window, can enter the monochromator and eventually the hollow-core fiber through higher order reflections and scattering inside of the monochromator.} Additional filtering either at \qty{4}{\kelvin} or at \qty{240}{\milli\kelvin} in the SNSPD housing could greatly reduce these background counts. The intrinsic dark count rates of the SNSPDs cannot be characterized as they are dominated by the blackbody radiation effects and relaxation oscillations at higher bias currents than shown in Fig.~\ref{PCR_360nm_130mV}. 

Figure~\ref{PCR_360nm_130mV}(b) shows measurement results of a \qty{200}{\nano\meter}-wide SNSPD measured in the cryostat with direct flood illumination with a thermal light source through bandpass filter stacks for longer wavelengths between \qty{10}{\micro\meter} and \qty{29}{\micro\meter}. The SNSPD shows saturated internal detection efficiency for all measured wavelengths. The slope in the plateau regime and the BCR are more prominent in these measurements. The reason might be the heating of the filter stack and other components in the cryostat due to the thermal light source. These components at elevated temperatures act as blackbody sources. Also, the bias current for this narrower SNSPD is much smaller, which requires a much lower trigger level. This increases the false counts due to electric noise, which can be seen as a constantly high BCR in Fig.~\ref{PCR_360nm_130mV}(b). 

Both measurements are limited by the cut-off bias current below which the amplitudes of the output pulses are below the trigger level and are therefore not counted. This impairs the shape of the photon count rate curve for the \qty{4}{\micro\meter} measurement in Fig.~\ref{PCR_360nm_130mV}(a) and the \qty{10}{\micro\meter} measurement in Fig.~\ref{PCR_360nm_130mV}(b). This could be improved by lower noise amplification and electronics that allow for lower trigger levels.

\begin{table*}
\caption{\label{tab:table1}\added{Comparison between WSi meander (\hbox{\qtyproduct{11 x 10}{\micro\meter}} area) taken from Taylor et. al. \cite{taylor_low-noise_2023} and WGe meander (\hbox{\qtyproduct{16 x 16}{\micro\meter}} area) measured in this work. The plateau width at \hbox{\qty{15}{\micro\meter}} wavelength is determined by $(I_{sw}-I(0.8))/I_{sw}$, where $I(0.8)$ is the bias current at which the normalized count rate is 0.8.}}
\renewcommand{\arraystretch}{1.5}
\begin{ruledtabular}
\begin{tabular}{@{}lcccccc@{}}
 &T$_c$ (K)& \thead{Film \\Thickness (nm)} & \thead{Operating \\Temperature (K)} & I$_{sw}$ (\unit{\micro\A}) & \thead{Nanowire \\Width (nm)} & \thead{Normalized Plateau \\Width (\qty{15}{\micro\meter} wavelength)}\\ \hline
 Taylor et. al. (WSi) \cite{taylor_low-noise_2023} & $1.3$ & $3$ & $0.25$ & $0.25$ & $80$ & $0.44$ \\
 This work (WGe) & $0.9$ & $8$ & $0.24$ & $0.85$ & $200$ & $0.41$ \\
\end{tabular}
\end{ruledtabular}
\end{table*}

\section{Conclusion}

In this work, we have developed WGe SNSPDs capable of single-photon detection at mid-IR wavelengths of up to \qty{29}{\micro\meter}. We have performed a comprehensive study of the WGe film composition and thickness, and its effects on the material properties and the performance of WGe SNSPDs. The fact that relatively wide nanowires on the order of \qty{400}{\nano\meter} can be used to obtain saturated internal efficiency at a wavelength of \qty{29}{\micro\meter} implies that WGe could be used to fabricate mid-infrared detectors over large areas with excellent yield. By comparison, nanowires fabricated from WSi must be on the order of \qty{100}{\nano\meter} or less to obtain saturated internal efficiency at \qty{10}{\micro\meter}\cite{Verma.2021, taylor_low-noise_2023}. Such narrow wires are difficult to yield over large areas. Furthermore, the WGe films used here were \qty{8}{\nano\meter} thick compared to previous work with WSi films which were less than \qty{3}{\nano\meter} thick\cite{Verma.2021, taylor_low-noise_2023}.\added{ To better illustrate the advantages of WGe compared to WSi, Table~\ref{tab:table1} compares the properties of a WSi nanowire meander from Taylor et. al.\cite{taylor_low-noise_2023} and the properties of a WGe nanowire meander from this work.} 

The use of thicker films that is possible with WGe should dramatically enhance yield over large areas and enable operation of these detectors at higher bias currents. Higher bias currents result in reduction in jitter which is important for potential applications in LIDAR. In multiplexing schemes such as the thermally-coupled imager (TCI) where Joule heating is used to trigger a transmission line bus, higher bias currents also result in higher Joule heating in proportion to the square of the current, making the use of such architectures feasible even in the mid-infrared with WGe. This shows that WGe may be a promising new SNSPD material for the fabrication of large-scale mid-infrared cameras that could be competitive with existing technologies such as MCT and BIB cameras. Such cameras would have the advantages of single-photon sensitivity and significantly lower noise in comparison to these technologies and could enable new applications in remote sensing, biological spectroscopy, chemistry, and astrophysics.  

\begin{acknowledgments}
This research was funded by NIST (https://ror.org/05xpvk416). Parts of this research were performed at the Jet Propulsion Laboratory, California Institute of Technology, under contract with the National Aeronautics and Space Administration (NASA—80NM0018D0004). Support for this work was provided in part by the DARPA DSO Invisible Headlights program and by the ERC IRASTRO program. DK thanks the EPSRC and SFI Centre for Doctoral Training in Photonic Integration for Advanced Data Storage (CDT-PIADS EP/S023321/1) and RHH thanks the UK Engineering and Physical Sciences Research Council (EP/S026428/1, EP/T00097X/1) for support. We thank Leaf Swordy and Katie David for thoughtful comments on the manuscript. We thank Chris Yung and Nathan Tomlin for assistance with the sputtering system used in this work. We thank the mechanical workshop and scientific staff at the Max Planck Institute for Multidisciplinary Sciences for help in the development of a cryogenically cooled infrared monochromator. 
\end{acknowledgments}

\section*{Author Declarations}
\subsection*{Conflict of Interest}
The authors have no conflicts to disclose.

\subsection*{Author Contributions} 
\textbf{Benedikt Hampel}:
Conceptualization (equal),
Data curation (lead),
Formal analysis (equal),
Investigation (lead),
Methodology (equal),
Resources (supporting),
Software (lead),
Validation (equal),
Visualization (lead),
Writing - original draft (equal),
Writing - review \& editing (lead)

\textbf{Daniel Kuznesof}:
Conceptualization (equal),
Data curation (supporting),
Formal analysis (equal),
Investigation (supporting),
Methodology (equal),
Validation (equal),
Writing - original draft (equal),
Writing - review \& editing (supporting)

\textbf{Andrew~S.~Mueller}:
Data curation (supporting),
Formal analysis (supporting),
Investigation (supporting),
Resources (supporting)

\textbf{Sahil R. Patel}:
Data curation (supporting),
Formal analysis (supporting),
Investigation (supporting),
Resources (supporting)

\textbf{Robert H. Hadfield}:
Funding acquisition (supporting),
Supervision (supporting)

\textbf{Emma~E.~Wollman}:
Funding acquisition (supporting),
Resources (supporting),
Supervision (supporting)

\textbf{Matthew D. Shaw}:
Funding acquisition (supporting),
Resources (supporting),
Supervision (supporting)

\textbf{Dirk Schwarzer}:
Funding acquisition (supporting),
Resources (supporting)

\textbf{Alec M. Wodtke}:
Funding acquisition (supporting),
Resources (supporting),
Writing - review \& editing (supporting)

\textbf{Khalid Hossain}:
Data curation (supporting),
Formal analysis (supporting),
Investigation (supporting),
Resources (supporting)

\textbf{Allison V. Mis}:
Data curation (supporting),
Formal analysis (supporting),
Investigation (supporting),
Resources (supporting)

\textbf{Alexana~Roshko}:
Data curation (supporting),
Formal analysis (supporting),
Investigation (supporting),
Resources (supporting),
Supervision (supporting)

\textbf{Richard P. Mirin}:
Funding acquisition (supporting),
Project administration (supporting),
Resources (supporting),
Supervision (supporting)

\textbf{Sae Woo Nam}:
Funding acquisition (supporting),
Project administration (supporting),
Resources (supporting),
Supervision (supporting)

\textbf{Martin J. Stevens}:
Funding acquisition (supporting),
Project administration (supporting),
Resources (supporting),
Supervision (supporting)

\textbf{Varun B. Verma}:
Conceptualization (equal),
Data curation (supporting),
Formal analysis (equal),
Funding acquisition (lead),
Investigation (supporting),
Methodology (equal),
Project administration (lead),
Resources (lead),
Supervision (lead),
Validation (equal),
Writing - original draft (supporting),
Writing - review \& editing (supporting)

\section*{Data Availability Statement}

The data that support the findings of this study are openly available in the NIST Science Data Portal at https://doi.org/10.18434/mds2-4009, Ref.~\cite{Hampel.2025}.

\appendix

\bibliography{WGe_SNSPDs}

\end{document}